\documentclass{PoS}

\title{Camera calibration strategy of the SST-1M prototype of the Cherenokov Telescope Array}

\ShortTitle{SST-1M Camera Calibration Strategy}

\author{\speaker{E. Prandini}$^{a}$, M. Heller$^{b}$, E. Lyard$^{a}$, E. Jr. Schioppa$^{b}$, and A. Neronov$^{a}$\\
        E-mail: \email{elisa.prandini@unige.ch} \\
W.~Bilnik$^{k}$,
J. B\l{}ocki$^{c}$,
L.~.Bogacz$^{m}$,
T~.Bulik$^{d}$,
F.~Cadoux$^{b}$,
A.~Christov$^{b}$,
M.~Cury{\l}o$^{c}$,
D.~della Volpe$^{b}$,
M.~Dyrda$^{c}$,
Y.~Favre$^{b}$,
A.~Frankowski$^{g}$,
\L{}. Grudniki$^{c}$,
M.~Grudzi{\'n}ska$^{d}$,
B.~Id{\'z}kowski$^{e}$,
M.~Jamrozy$^{e}$,
M.~Janiak$^{g}$,
J.~Kasperek$^{k}$,
K.~Lalik$^{k}$,
E.~Mach$^{c}$,
D.~Mandat$^{l}$,
A.~Marsza{\l}ek$^{c,e}$,
J.~Micha{\l}owski$^{c}$,
R.~Moderski$^{g}$,
T.~Montaruli$^{b}$,
J.~Niemiec$^{c}$,
M.~Ostrowski$^{e}$,
P.~Pa{\'s}ko$^{f}$,
M.~Pech$^{l}$,
A.~Porcelli$^{b}$,
P.~Rajda$^{k}$,
M.~Rameez$^{b}$,
P.~Schovanek$^{l}$,
K.~Seweryn$^{f}$, 
K.~Skowron$^{c}$,
V.~Sliusar$^{j}$,
M.~Sowi{\'n}ski$^{c}$,
{\L}.~Stawarz$^{e}$,
M.~Stodulska$^{e}$,
M.~Stodulski$^{c}$,
S.~Toscano$^{b,n}$, 
I.~Troyano Pujadas$^{b}$, 
R.~Walter$^{a}$,
M.~Wi{\c e}cek$^{k}$,
A.~Zagda\'{n}ski$^{e}$,
K.~Zi{\c e}tara$^{e}$,
P.~{\.Z}ychowski$^{c}$  for the CTA Consortium \\
\footnotesize{
a. ISDC, Observatoire de Gen\`eve, Universit\'e de Gen\`eve, 1290 Versoix, Switzerland.\\
b. D\'epartment de physique nucleaire et corpusculaire, Universit\'e de Gen\`eve, CH-1205 Switzerland.\\
c. Instytut Fizyki J{\c a}drowej im. H. Niewodnicza{\'n}skiego Polskiej Akademii Nauk,  31-342 Krak{\'o}w, Poland.\\
d. Astronomical Observatory, University of Warsaw, Al. Ujazdowskie 4, 00-478 Warsaw, Poland\\
e. Astronomical Observatory, Jagiellonian University, ul. Orla 171, 30-244, Krak{\'o}w, Poland.\\
f. Centrum Bada{\'n} Kosmicznych Polskiej Akademii Nauk,  18a Bartycka str., 00-716 Warsaw, Poland.\\
g. Nicolaus Copernicus Astronomical Center, Polish Academy of Sciences,  Warsaw, Poland.\\
j. Astronomical Observatory, Taras Shevchenko Nat. University of Kyiv, Observatorna str., 3, Kyiv, Ukraine.\\
k. AGH University of Science and Technology, al.Mickiewicza 30, Krak{\'o}w, Poland,\\
l. Institute of Physics of the Czech Academy of Sciences, Prague, Czech Republic.\\
m. Department of Information Technologies, Jagiellonian University, 30-348 Krak{\'o}w, Poland.\\
n. Vrije Universiteit Brussels, Pleinlaan 2 1050 Brussels, Belgium.\\}
}
  
\abstract{The SST-1M telescope is one of the prototypes under construction proposed to be part of the future Cherenkov Telescope Array. It uses a standard Davis-Cotton design for the optics and telescope structure, with a dish diameter of 4 meters and a large field-of-view of 9 degrees.
  The innovative camera design is composed of a photo-detection plane with 1296 pixels including entrance window, light concentrators, Silicon Photomultipliers (SiPMs), and pre-amplifier stages together with a fully digital readout and trigger electronics, DigiCam.
  
  In this contribution we give a general description of the analysis chain designed for the SST-1M prototype. In particular we focus on the calibration strategy used to convert the SiPM signals registered by DigiCam to the quantities needed for Cherenkov image analysis. The calibration is based on an online feedback system to stabilize the gain of the SiPMs, as well as  dedicated events (dark count, pedestal, and light flasher events) to be taken during the normal operation of the prototype.}

\FullConference{The 34th International Cosmic Ray Conference,\\
		30 July- 6 August, 2015\\
		The Hague, The Netherlands}

\begin{document}

\section{Introduction}
The Cherenkov Telescope Array (CTA) will be the first open access observatory  for very high-energy (VHE) gamma rays. Two arrays of telescopes in a northern and a southern site will allow to cover the full sky.  The two CTA arrays will be composed of more than a hundred Imaging Atmospheric Cherenkov Telescopes (IACTs) of three different sizes,  among  which the small size telescopes (SSTs) are for observing the highest energies.  The construction of the first telescopes prototypes started in 2014. 

In order to study the VHE gamma ray sky,  IACTs observe the faint Cherenkov light emitted by extensive air showers induced in the atmosphere by incident cosmic-ray particles and gamma-rays.

If the telescope is in the footprint of the cherenkov light pool on the ground, each
 shower creates one image in the camera. 
The study of these images allows the reconstruction of the energy and incoming direction
of the primary gamma-ray. A crucial part of the data analysis is represented by the
rejection of background events, i.e. triggered images induced by cosmic rays (mainly protons).

In this contribution we  give a general description of the analysis chain designed for 
the SST-1M prototype. In particular we focus on the calibration strategy used to convert 
the  signals registered by the camera to the quantities needed for Cherenkov image analysis. 

\section{The SST-1M Project}
The SST-1M telescope is one of the CTA prototypes under construction in Krakow by a Polish and Swiss consortium, 
and is the only SST prototypes designed with a single mirror. 
It uses a standard Davis-Cotton design for the optics and telescope structure, 
with a dish diameter of 4 meters and a large field-of-view of 9 degrees \cite{bib:TM}. 


The SST-1M consortium adopted the novel approach of having the photo-detection plane (PDP) and
the digital signal acquisition/readout electronics (DigiCam) as two physically separated
entities, both hosted in the camera. The PDP is developed by the University of Geneva, while the DigiCam digital signal 
acquisition and readout systems are developed by the team of the Jagiellonian University and
AGH University of Science and Technology, Krakow.

\subsection{The Photo-Detection Plane}
The PDP is composed of the following elements (from outer to inner part of the camera): light funnels, light detectors, preamplifier boards, and slow control boards.
The main novelty of the PDP concept of the SST-1M prototype is that silicon photomultipliers (SiPMs) are used as photon detectors.
This choice was driven by the very successful results obtained with the FACT telescope \cite{bib:FACT}, the first IACT with a SiPM camera. 

The SST-1M camera  is  shaped into an hexagon which is $\sim$1~m flat-to-flat 
wide and about 60~cm thick. It is equipped with 1296 hexagonal SiPM detectors of 5.5~mm side and 10~mm flat-to-flat. 
In front of each detector, a hollow light funnel effectively increases the pixel size \cite{bib:cones}.
The final angular pixel size of 0.24 degrees (2.4~cm) was selected based on the  angular resolution 
required by the CTA consortium.
The PDP will therefore be a mosaic of hollow cones coupled to the sensor, thus reducing the dead
space and cut off background light not coming from the mirror.

The amplification of the SiPMs signals is performed by the Preamplifier board. The board also routes the signal to the slow control board.

Each SiPM detector of the SST-1M camera has a temperature sensor. This is fundamental for the performance of the camera and the calibration of the signal, as the sensor gain is strongly temperature dependent. In particular, the gain depends on the overvoltage, i.e. the difference between its operational voltage and the  breakdown voltage (which is temperature dependent). Given a certain level of Night Sky Background (NSB), during normal operations the temperature of the sensor, and hence its breakdown voltage, will change. The gain can, however, be kept constant by adjusting the operational voltage so that the overvoltage is constant.
This task is done inside the camera by the slow control board, which reads the temperature of each sensor and adjust the voltage accordingly. In addition this board routes the amplified signal to the DigiCam. 

\subsection{DigiCam}
The design of the SST-1M digital readout and trigger systems 
descends from the need
of a compact, high performing electronics, which at the same time has to be as standard as
possible and easily mass producible \cite{bib:digicam}.
The whole digital subsystem of the camera  consists of three mini-crates each containing
the same set of boards and serving one third of the PDP area (i.e., 432 pixels). 
The analog signals from the PDP will be transferred to the digital subsystem, 
where they are first digitized by the Flash Analog-to-Digital Converter (FADC) board.

The {\it FADC board} main tasks are:  digitize the analog signals of
the PDP,  pre-process digitized signals and store them in digital ring buffers, calculate
the first level trigger (L0) signals and send them continuously to the trigger board.
 The L0 trigger is based on a 3-pixel set (triplets) allowing for multiplicity and/or 
summed amplitude triggers (clipped sum trigger). 

The trigger board is therefore  in charge of receiving the trigger signals
from all the FADC boards within the crate and from the neighbouring channels in the
remaining sectors and calculating the second level trigger (L1) signal. L1 trigger
is based on the spatial distribution of L0 triggers.
The decision is then sent over the whole digital electronics.  The trigger board is also in charge of collecting
the event data from all the FADC boards within the crate and send them to the central
acquisition system (through the master trigger card).

\subsection{The camera server}
The camera server is not part of the camera itself, and will
likely be located in the common server room at the site. Its main
task is to receive the locally trigger data from the camera and send the triggered signals
 to the data acquisition system (DAQ).

\section{Data analysis Pipeline for the SST-1M prototype}
The data analysis chain for the SST-1M prototype telescope is currently under construction. 
A general scheme of the analysis steps based on the experience of running IACTs is shown
in Figure 1 and includes:
\begin{description}
\item[Calibration and Image Cleaning:]  raw data (DL0) are processed into calibrated data (DL1).
The first step of the analysis is represented by the {\it calibration for image reconstruction or camera calibration}, that concerns the relative  calibration of pixels in a camera, and the estimation of the level of statistical fluctuations of pixel signals,  as well as relative time offsets between pixels. Dark counts and flasher events are needed as input data.
In addition to the camera calibration,  the {\it pointing calibration} is also done. It concerns the determination of the effective optical axis of the telescope, in absolute (sky) coordinates, and the determination of the conversion between camera and sky coordinates. The input data are the starguide data.
An accurate atmospheric calibration can be additionally performed in case that atmospheric monitoring data are collected.
After this step, the calibrated image is cleaned, in the sense that only the pixels forming the image, i.e. surviving pre-defined cuts, survive. 
\item[Image Parameters Reconstruction:]  DL1 are processed into reconstructed data (DL2). The pixels surviving the cleaning form an image that, in case of gamma-like events, is expected to be an ellipse in the camera plane which is parametrized by several  parameters, such as Hillas  and timing parameters. For some of the reconstructed parameters, such as energy and direction, look-up tables and/or Monte Carlo data pre-processed in an analogue pipeline are used.
\item[Data Reduction:]  reduced data (DL3) are produced. During this step, only selected events (i.e. gamma-like) are extracted from the overall event list. Monte Carlo data could be used at this step to estimate additional image parameters.
\item[Scientific Analysis:] science data (DL4) are produced (with the so-called {\it ctools}). During this step, high-level binned data products like spectra, skymaps, or lightcurves are generated. Input for the process is the instrument response function.  
\end{description}

\begin{figure}[htp]
  \centering
  \includegraphics[width=5.0in]{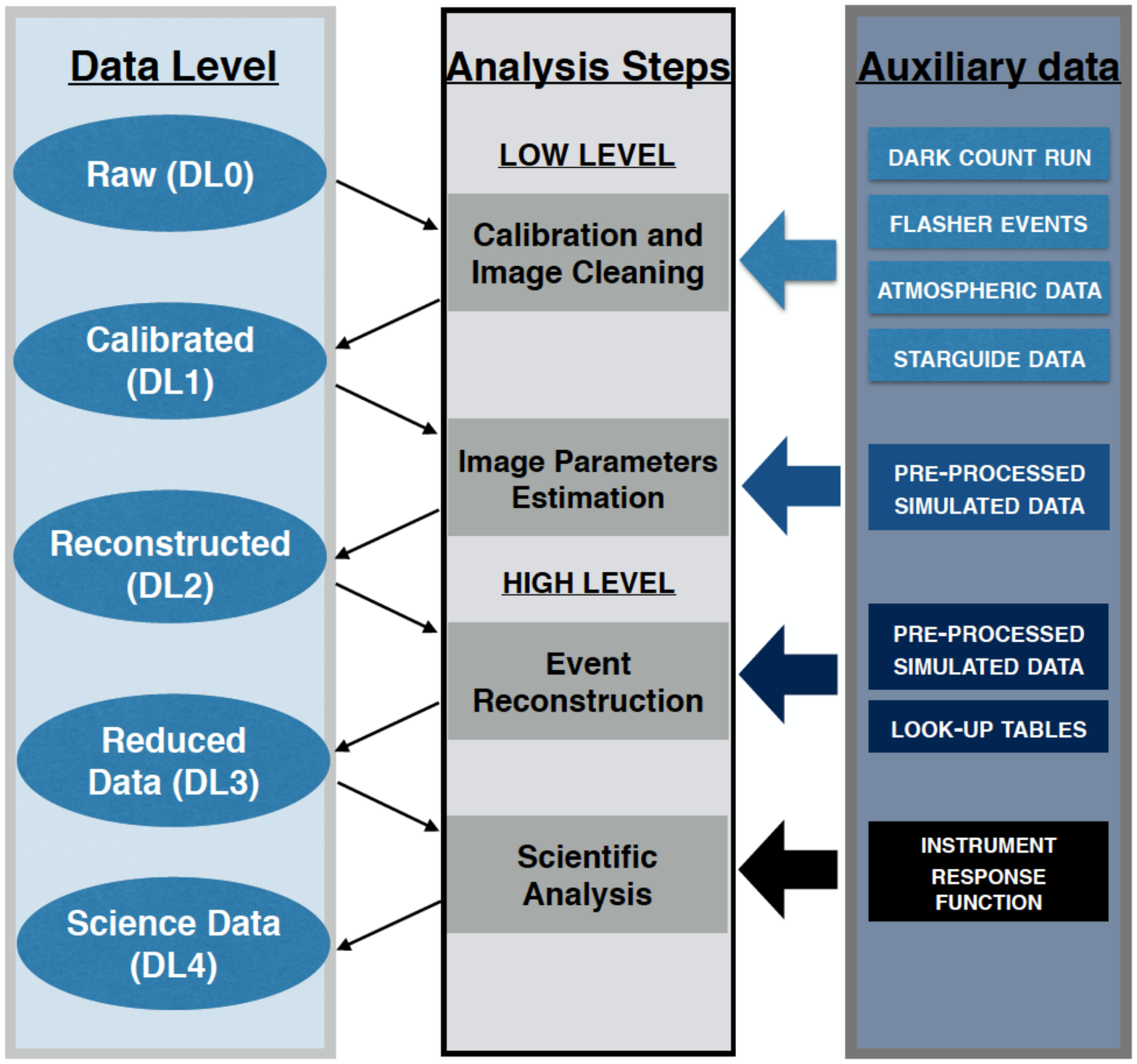}
  \caption{A scheme of the data analysis pipeline and scientific analysis foreseen for the SST-1M prototype.}
  \label{sst1m_pipeline}
 \end{figure}

\section{Camera Calibration}

The first step of the analysis chain is the camera calibration. 
As described in the previous section, it
concerns the relative calibration of pixels in the camera, i.e. 
the signal extraction and conversion into
physical units, and the estimation of the level of statistical fluctuations of pixel signals.

\subsection{Waveform extraction}
For the SST-1M prototype, raw DL0 data are the signal waveforms acquired with a sampling period of 4~ns. Each waveform is made of 20 samples. 
In order to extract the maximum amplitude and the integral of the waveform, the baseline must be subtracted. 
A time window of 20 samples is acquired before the signal in the so-called pedestal events, which is possible thanks to the ring buffers.
This window is used to calculate the baseline level and its fluctuations. 
The baseline level is subtracted from the waveform allowing the extraction of the maximum pulse amplitude and its integral.
However the calibrated DL1 data must be expressed in term of photo-electrons (pe). Therefore the conversion factors from the maximum amplitude of the pulse to pe and from the integral of the pulse to pe must be known.

\subsection{Conversion factor}
\begin{figure}
\begin{center}
\includegraphics[width=0.8\textwidth]{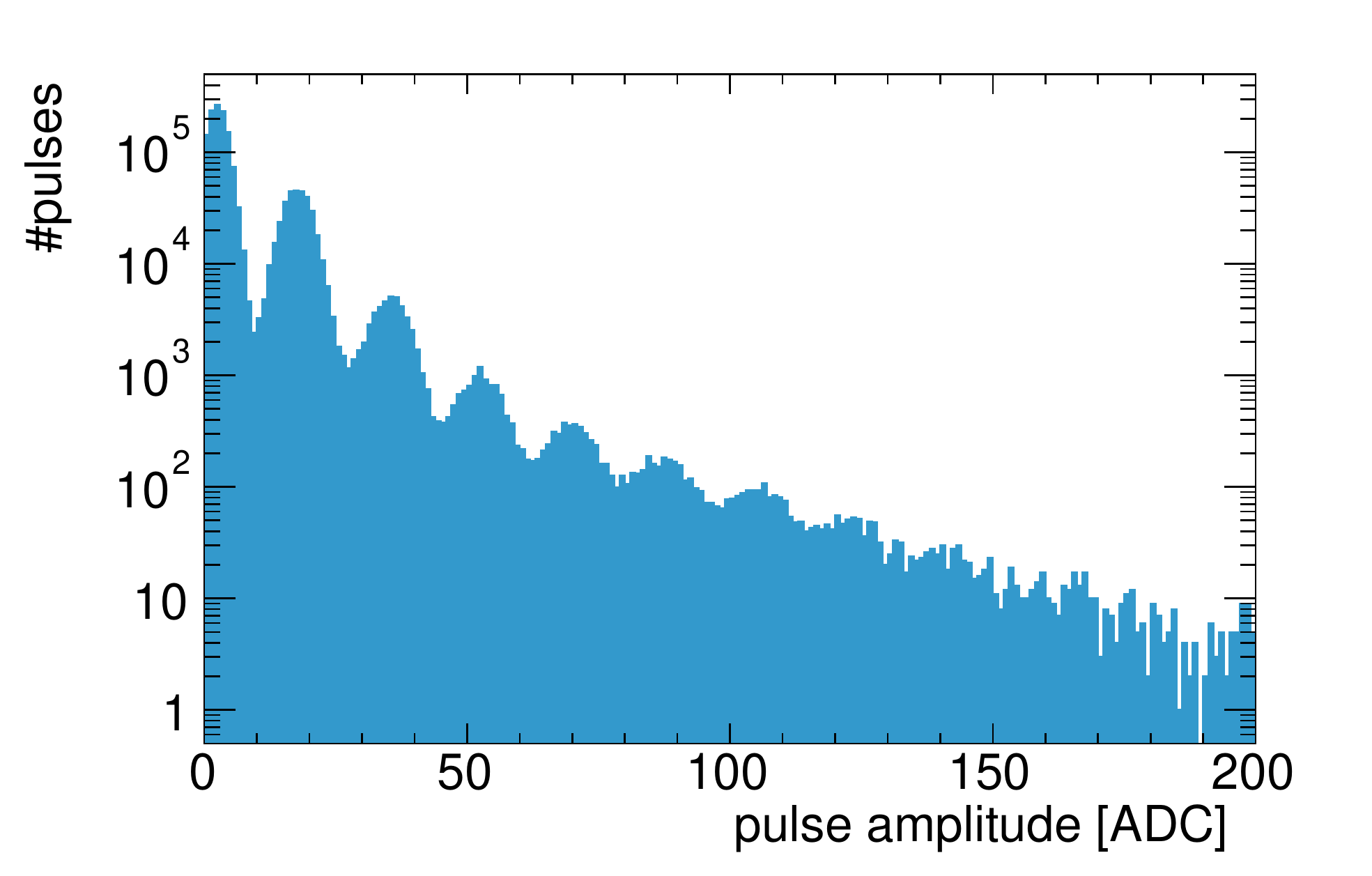}
\caption{Amplitude distribution of all pulses detected during a dark count run of $10^5$ acquisition windows of 2~$\mu$s each.}
\label{fig:SPEspectrum}
\end{center}
\end{figure}

The advantage of SiPM compared to standard photomultiplier tubes is that they are  better to distinguish single photons. However this is valid only for light levels typically less than 10 pe per pulse in average.
The method to extract the conversion factor, also called gain, is to calculate the equivalent signal (maximum amplitude or pulse integral) for a single photon.To do so, the SST-1M plans to have two types of low light level runs:
\begin{description}
\item[dark count runs] are acquired when the lid is closed such that no external light is seen by the SiPM and that  the only signal generated is from thermally excited electrons.
Figure \ref{fig:SPEspectrum} shows the typical results of such runs. The distribution shown is the reconstructed maximum amplitude for all the recorded waveforms. While the first peak is the pedestal, the following ones are the single and multiple pe contributions. The distance between the peaks allow to extract the convertion factor.
\item[flasher runs] are acquired during the physics data taking. In order to avoid that these events overlap with physics event, the light is flashed at a lower rate than the expected physics trigger rate. The light produced by the flasher must be uniform and should have an average of few pe per pixel. This allows to monitor continuously the gain of the electronics chain.
\end{description}

In addition to the conversion factor, the dark count runs allow to measure the optical cross talk. 
This cross talk introduce an excess in the number of detected photons. Its value depends on the overvoltage applied to the pixel and can only be measured during dark count runs. 
Indeed, the multiple photon events can be related to either coincidence between dark count events or, and most likely, to cross talk. 
Once the contribution from coincidences is subtracted the cross talk value is extracted and taken into account in the conversion factor calculation.

As mentioned in previous sections, the temperature compensation loop allows to maintain a constant gain along the runs. 
However, it was predicted by electronics simulation, and measured in the laboratory that the convertion factor depends on the NSB level.

\subsection{Conversion factor dependency on NSB level}
The activity on the sensors caused by the NSB photons induces a voltage drop in the biasing circuit of the pixel resulting in a decrease of the overvoltage. As explained in previous sections the consequence is a decrease in the pixel gain and therefore in the conversion factor. This effect has been characterized in laboratory for various NSB levels and a parametrization was extracted.
The NSB level can be determined from the baseline level (DC coupling) and its fluctuations which are both measured with the pedestal events. 
A reference baseline measurement is obtained during the dark count runs. In doing so, both the electronics noise and the dark count events contributions are subtracted leaving only the pure NSB contribution.
Once this is known, the parametrization provides a correction coefficient and allows us to 
scale accordingly the conversion factor.
The flasher runs will be used to monitor the conversion factor along the data taking.

For the camera calibration, other parameters will be measured with a lower periodicity ($\sim$ once per year). They are detailed in \cite{bib:MD} and can be divided into four main topics:  the photo detection efficiency dependencies, the pixel photon conversion efficiency, the flat-fielding corrections, and the signal integration.

\subsection{Calibration Hardware}

The measurements needed to extract the different parametrization (cross talk, gain vs NSB, gain uniformity) will be tested with a dedicated setup in the laboratory. 
The setup used \cite{bib:EJS} allows to calibrate four modules (i.e. group of 12 pixels) at a time using DigiCam \cite{bib:Rajda} in both continuous and pulsed light mode. 
Optical fibers are brought in front of each pixel and illuminate not only the sensor but also the light guides from their entrance point. Therefore also the optical efficiency of each pixel is measured and calibrated.

Concerning the apparatus required for the on-site calibration, dedicated hardware needs to be developed \cite{bib:MD}. 
The SST-1M group will use a high precision external flasher unit developed by another camera group of CTA (\cite{bib:Nectar}, \cite{bib:GCT}). This flasher will be used for the so-called flasher runs.

Concerning all the yearly calibrations, the SST-1M group will favour external and movable systems in order not to have to unmount and transport the camera. Doing so, the camera does not need to be taken out of its basket and unnecessary mechanical stress can be avoided.

\begin{figure}
\begin{center}
\includegraphics[width=0.45\textwidth]{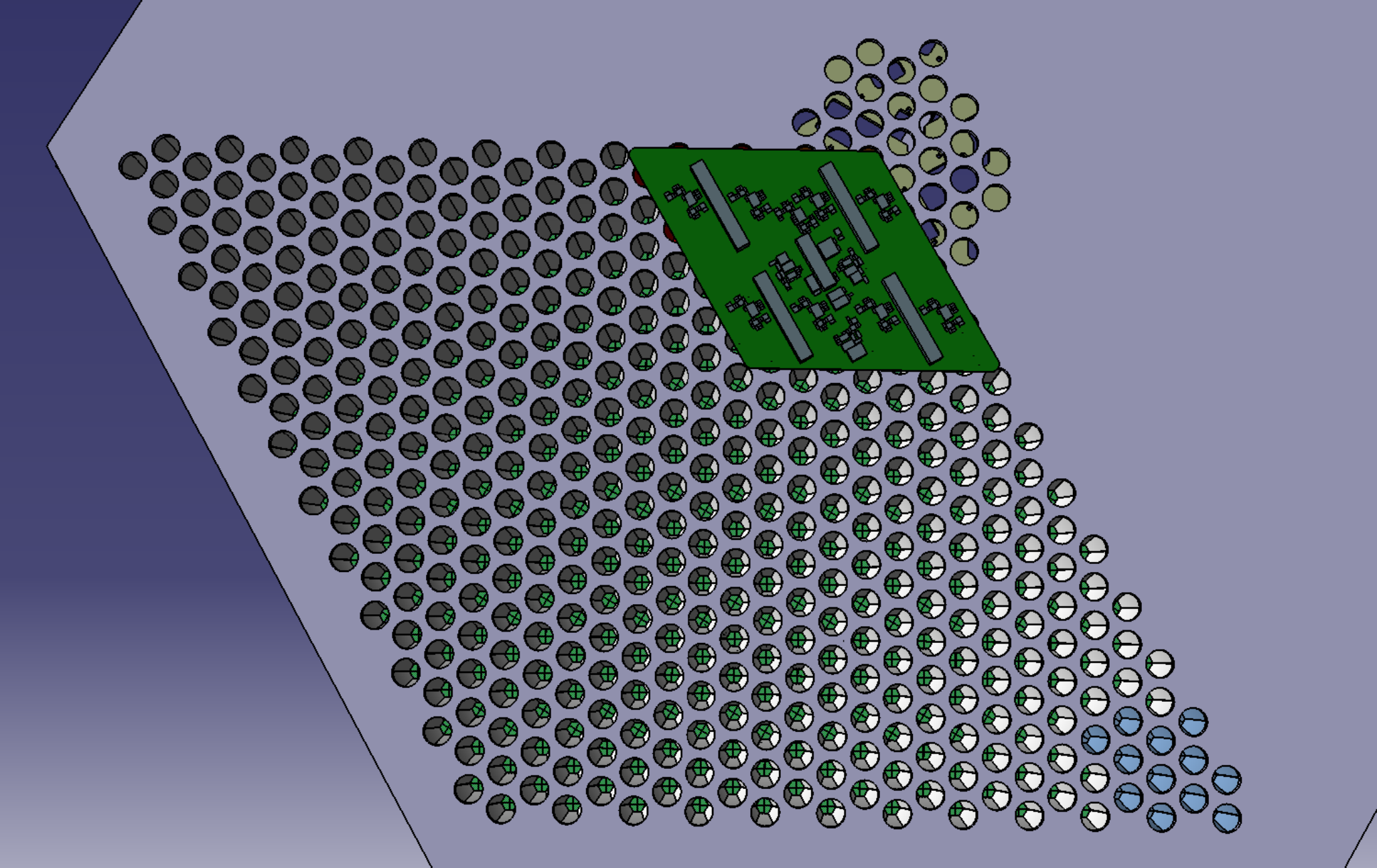}\hspace{0.025\textwidth}
\includegraphics[width=0.417\textwidth]{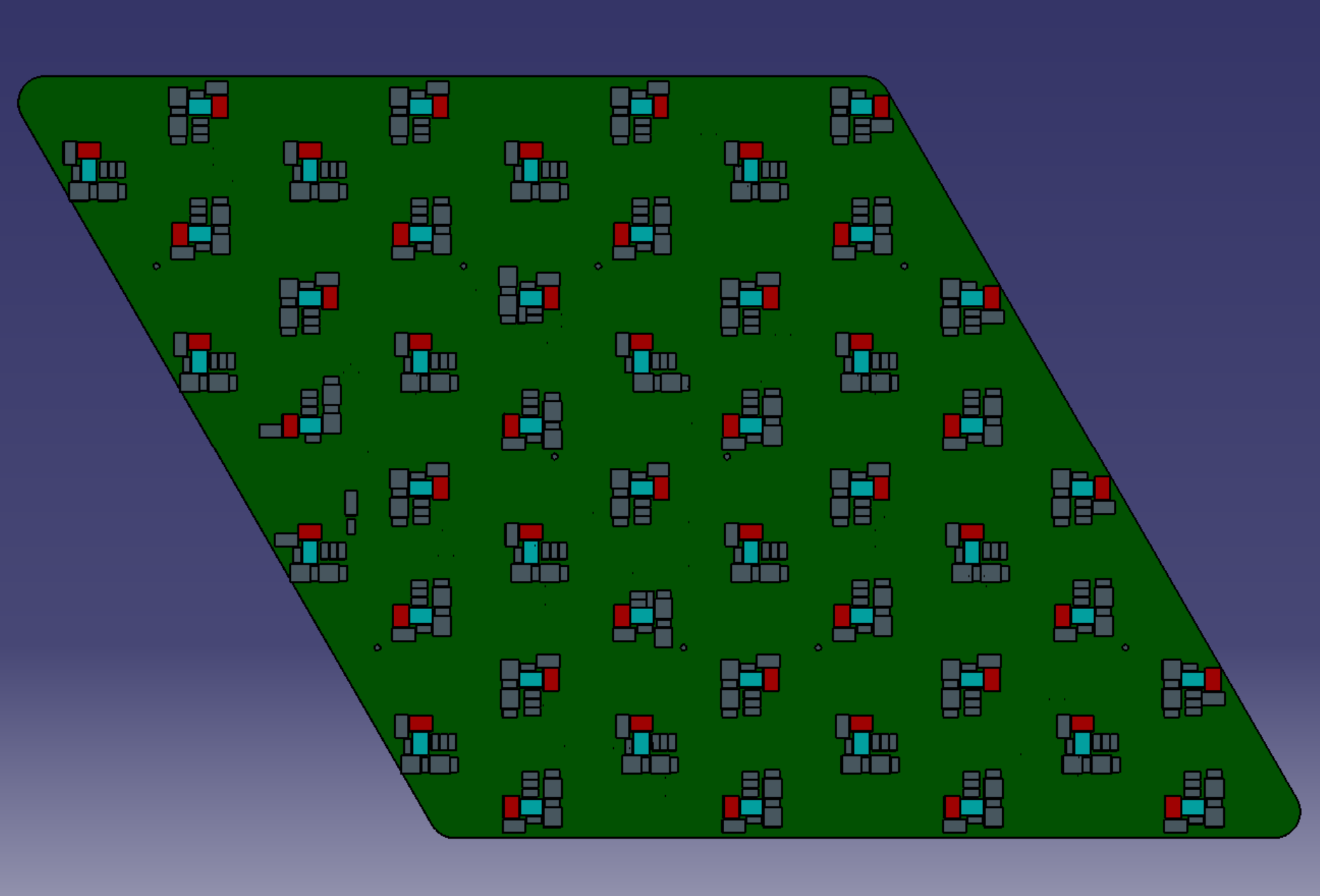}
\caption{Left: Overview of the the camera test setup with one LED flasher module mounted. Right: Detailed design drawing of an LED flasher module. The blue component is the flashed LED while the red one is the continuous LED.}
\label{fig:CTS}
\end{center}
\end{figure}

In addition to common calibration devices, the SST-1M group has designed an innovative setup in order to scrutinize the entire acquisition chain. This setup  will be used in laboratory and is foreseen to be moved  on-site, for yearly calibration.
On top of the camera window a mechanical support covering one third of the camera and its central part will hold two 420~nm LEDs per pixel. One will be pulsed while the other can deliver a continuous light in order to emulate the NSB. LEDs can be individually switched on and their light level adjusted in groups of three. Design drawings are shown on figure \ref{fig:CTS}.
This will allow not only to probe possible cabling problems but also to train the different trigger logics as complex light patterns can be created. 

\section{Conclusions}
In this contribution, the main steps of the data analysis foreseen for the SST-1M prototype have been outlined. In particular, we have described in detail the camera calibration of the prototype, designed to fulfill the CTA requirements, and reported the needed calibration hardware. 

In order to calibrate the signal, the maximum amplitude and the integral of the waveform is extracted from the raw DL0 data, pedestal subtracted. 
Two additional dedicated events will be taken for the calibration. A dark count run will be performed at the beginning of the observations for the conversion factor estimation and the cross talk correction. Moreover, flasher events needed for the monitoring of the conversion factor will be taken interleaved with physical events.

\section*{Acknowledgements}
We gratefully acknowledge support from the agencies and organizations listed under Funding Agencies at this website: http://www.cta-observatory.org/.
EP gratefully acknowledges the financial support  of the Marie Heim-Vogtlin grant of the Swiss National Science Foundation.

\end{document}